**Direct measurement of electron intervalley relaxation in a Si/SiGe quantum dot**

Nicholas E. Penthorn[1], Joshua S. Schoenfield[1], Lisa F. Edge[2], and HongWen Jiang[1]

[1]Department of Physics & Astronomy, University of California, Los Angeles, California 90095, USA
[2]HRL Laboratories, LLC, Malibu, California 90265, USA

**Abstract**

The presence of non-degenerate valley states in silicon can drastically affect electron dynamics in silicon-based heterostructures, leading to electron spin relaxation and spin-valley coupling. In the context of solid-state spin qubits, it is important to understand the interplay between spin and valley degrees of freedom to avoid or alleviate these decoherence mechanisms. Here we report the observation of relaxation from the excited valley state to the ground state in a Si/SiGe quantum dot, at zero magnetic field. Valley state read-out is aided by a valley-dependent tunneling effect, which we attribute to valley-orbit coupling. We find a long intervalley relaxation time of $12.0 \pm 0.3$ ms, a value that is unmodified when a magnetic field is applied. Furthermore, we compare our findings with the spin relaxation time and find that the spin-valley "hot spot" relaxation is roughly four times slower than intervalley relaxation, consistent with established theoretical predictions. The precision of this technique, adapted from electron spin read-out via energy-dependent tunneling, is an improvement over indirect valley relaxation measurements and could be a useful probe of valley physics in spin and valley qubit implementations.

## I. INTRODUCTION

Valley physics has been the subject of many theoretical and experimental studies in silicon-based quantum dot devices, largely in association with semiconductor quantum computing [1] [2] [3] [4]. Although the six valleys in the silicon conduction band are degenerate in the bulk crystal [5], electron confinement in heterostructures lifts the degeneracy and leads to two low-lying valley states, one of which is the ground state for electrons confined within the quantum dot potential [6] [7] [8] [9]. Out-of-plane confinement created from gate-generated electric fields and the quantum well interface, in addition to interface disorder, determines the magnitude of the energy splitting between the two valleys [10] [11] [12]. The presence of these states is generally considered to be a hinderance to the performance of electron spin qubits in silicon, in part because noise sources can couple to spin via the valley degree of freedom to give rise to spin relaxation [13] [14] [15]. Moreover, the spin qubit operation point must be carefully chosen to avoid the fast relaxation "hot spot" where the spin and valley states mix [13] [15] [16].

Several studies have found theoretically that spin relaxation in silicon is dominated by spin-valley hybridization effects at external fields used for spin quantum computing, $B_{\text{ext}} \approx 0.2 - 2$ T. In this regime, the (intravalley) spin relaxation rate depends on the intervalley relaxation rate $\Gamma_v$ as well as the proximity of the Zeeman level splitting to the valley splitting $E_{\text{VS}}$ [13]:

$$\Gamma_{\text{spin}} = \frac{\sqrt{\delta^2 + \Delta_a^2} + |\delta|}{2\sqrt{\delta^2 + \Delta_a^2}} \Gamma_v \tag{1}$$

In (1), $\delta = E_Z - E_{\text{VS}}$ is the difference between the Zeeman and valley splitting energies and $\Delta_a$ is the spin-valley coupling strength. Importantly, the intervalley relaxation rate $\Gamma_v$ has polynomial growth with the magnitude of the valley splitting, with Johnson noise contributing a linear dependence and the valley-phonon interaction adding a $E_{VS}^5$ dependence [15] [16] [14] [17]. In Si/SiO$_2$ quantum dots, $E_{VS}$ can range from 0.1 to 1 meV [18] [19] and $\Gamma_v$ is expected to be as fast as 100 MHz [13] [17]. Consequently, a direct measurement of intervalley relaxation in quantum dots has never been achieved, as the relaxation rate is just outside the bandwidth of state-of-the-art charge detection with rf reflectometry [20] [21] [22] and far outside the bandwidth of standard capacitive charge sensing measurements. However, the valley splitting in Si/SiGe-based quantum dots is almost universally smaller than their Si/SiO$_2$ counterparts [23] [24] [25] [26] [27] and could result in a relaxation rate slow enough to capture in real-time. This work utilizes a small valley splitting of order 10 µeV found in a Si/SiGe quantum dot device to directly observe intervalley relaxation with no applied magnetic fields and show that the measured relaxation time $T_1 = 1/\Gamma_v$ is consistent with indirect measurements extracted from spin relaxation in the vicinity of the spin-valley "hot spot". Robust read-out of the valley occupation despite the small energy splitting is aided by valley-dependent tunneling, possibly originating from valley-orbit coupling. We then show that intervalley relaxation and spin relaxation can be seen at the same magnetic field value when the quantum dot energy is modified slightly. Valley $T_1$ is found to be independent of magnetic field, implying a similarly field-independent valley splitting and negligible contributions from second-order relaxation processes.

## II. DEVICE CHARACTERIZATION

The device under study was a gate-defined double quantum dot fabricated on a Si/SiGe wafer known from an earlier work to have small valley splitting [28] (Fig. 1(a)). Positive voltage on a global accumulation gate creates a two-dimensional electron gas and voltages on local depletion gates define potential wells in the plane of the wafer. A single-electron transistor (SET) formed in the upper channel is used to detect changes in electron occupation of the two quantum dots, defined in the lower channel. For this work, inter-dot tunneling was suppressed by setting an appropriately negative voltage on barrier gate *M*, effectively creating two non-interacting charge boxes that are each coupled to separate electron reservoirs. The dots' energy can be controlled with voltages on plunger gates *L* and *R* and the electron tunneling rates into and out of each reservoir were controlled with barrier gate voltages *BL* and *BR*. Both quantum dot energies were tuned to the single electron limit, so that at any point during experiments there is either one or zero electrons in each dot (Fig. 1(b)). Dot occupation was confirmed by magnetospectroscopy [29]. We focus on the right dot since the valley splitting in the left dot was found previously to be around 20 µeV [28], too small to perform valley read-out with our system.

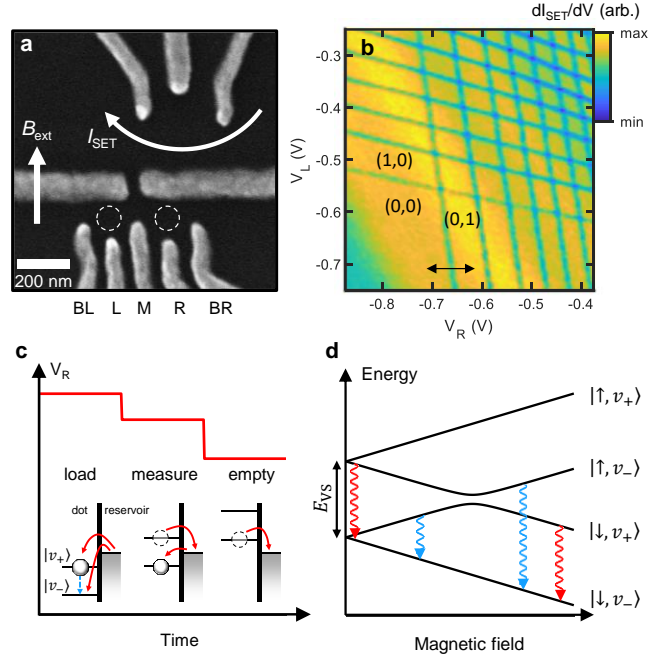

FIG. 1. Device characterization and measurement details. (a) Scanning electron micrograph of the double dot device. Arrows represent the direction of current through the single-electron transistor $I_{SET}$ and the applied magnetic field $B_{ext}$, where applicable. (b) Charge stability diagram of the double quantum dot with inter-dot tunneling suppressed. Indices (*i*, *j*) indicate the occupation of the (right, left) dot. Black arrow shows the location of voltage pulses on gate *R*. (c) Diagram of the three-stage pulse used to convert spin-valley state occupation to a SET current signal. During the "load" stage, an electron in the excited valley can relax to the ground state.

(d) Spin-valley levels as a function of in-plane magnetic field, including the relaxation mechanisms probed in this study: intervalley (red arrows) and spin (blue arrows).

III. MEASURING VALLEY STATE OCCUPATION

Conversion of single-electron valley state occupation to charge for SET readout was accomplished in real time using energy-dependent tunneling [30] [31] [32]. A three-stage voltage pulse modifies the dot energy as shown in Fig. 1(c). First an electron is loaded into the dot from the reservoir and held there for time $t_\text{wait}$. Assuming reservoir electrons occupy both valley states evenly, the loaded electron will occupy the excited valley or the ground valley with equal probability. In the measurement phase of the pulse, the dot energy is raised so that the valley states straddle the reservoir Fermi energy. If the loaded electron is in the excited valley at the beginning of this phase, it can tunnel into the reservoir and tunnel back into the dot's ground state, registering as an abrupt and transient increase in SET current. The principle of this operation is the same as the commonly-used technique for measuring single spin relaxation, except here the measurement takes place at zero applied magnetic field so that the Zeeman splitting is much smaller than the valley splitting and the readout level lies between valley states, not spin states (see Fig. 1(d) for relaxation processes that can be observed with this technique). In our device the valley splitting seen in both dots is comparable to thermal broadening, so we analyzed valley occupation with two thousand averages of the pulse sequence. When averaged, the random transient signal generated by an excited state electron becomes a "tunneling peak" that can be easily discriminated from thermally-driven random telegraph signal [33]. The SET current, which contains the electron response to the pulse and a comparable component that arises from the capacitive coupling of the pulsed gate to the SET, can be converted to a probability of the dot being unoccupied. This is accomplished by performing an identical three-stage pulse far enough away from the charge transition that no electron tunneling occurs. The resulting SET current contains only the capacitive coupling to the pulsed gate and can be subtracted from the original data to reveal the pure electron response to the pulse. Then, provided the "load" and "empty" stages of the pulse are long enough compared to the electron tunneling rates, the electron response can be scaled so that the unoccupied probability is zero at the end of the "load" stage and one at the end of the "empty" stage. Fig. 2(a) shows the probability of the right dot to be empty over the three-step pulse, with a tunneling peak appearing in the measurement phase. Remarkably, the tunnel-in and tunnel-out rates contributing to the valley tunneling peak are much lower than the rates associated with the electron loading and unloading pulse stages. Modelling the probability of dot occupation with coupled classical rate equations and taking thermal effects into account [34, 35], the slow electron hopping that comprises the tunneling peak can be fully explained by a dependence of the tunneling rate on the occupied state, temperature and the direction of tunneling.

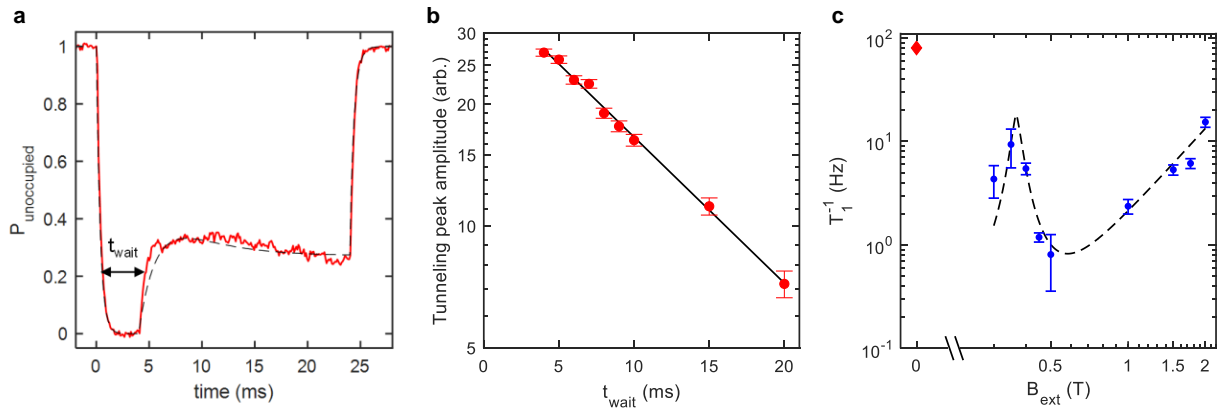

FIG. 2. Valley and spin relaxation. (a) Probability of the right dot being unoccupied during the pulse sequence with $t_{\text{wait}} = 4$ ms. The tunneling peak is visible during the read-out pulse stage indicating upper valley state occupation. Black dotted line: fit to rate equation model. (b) Tunneling peak amplitude as a function of time spent in the load stage $t_{\text{wait}}$. The trend follows exponential decay with valley $T_1 = 12.0 \pm 0.3$ ms. (c) Blue circles: spin relaxation rate at nonzero fields shows a hot spot at $B_{\text{ext}} = 0.37 \pm 0.03$ T where spin and valley states mix. Red diamond: intervalley relaxation rate $\Gamma_v = 1/T_1 = 83 \pm 2$ Hz at zero field. Error bar is smaller than the diamond and is not shown.

## IV. INTERVALLEY RELAXATION

Because any electron loaded into the dot's excited valley can decay to the ground state in the presence of intervalley relaxation, the amplitude of the tunneling peak will decrease exponentially as a function of $t_\text{wait}$ with a characteristic relaxation time $T_1$ as $e^{-t_\text{wait}/T_1}$. Fig. 2(b), which is the main result of this paper, shows the direct observation of intervalley relaxation with a valley $T_1$ of 12.0 ± 0.3 ms. As expected from the small valley splitting, this value is long when compared to previous estimates in Si/SiO$_2$ and Si/SiGe [13] [24]. Moreover, this result is the first precise measurement of intervalley relaxation in silicon quantum dots.

A direct comparison between spin and valley relaxation was made by repeating the T$_1$ measurement at nonzero applied magnetic field, in the limit where the Zeeman splitting is comparable to or larger than the valley splitting. Plotting the spin relaxation rate as a function of field reveals the well-studied hot spot where the Zeeman and valley levels mix (Fig. 2(c)). Relaxation at the spin-valley anti-crossing increases sharply and is expected to be on the order of the bare valley relaxation, shown by setting $\delta = 0$ in (1). We find a hot spot at 0.37 ± 0.03 T in the right dot, corresponding to a valley splitting of $E_\text{VS} = 43 \pm 4$ µeV, consistent with previous measurements obtained from Ramsey fringe experiments on the same device [28]. From a fitting of the relaxation curve, including phonon and Johnson noise contributions [14] [15] [16] [36], we find that the relaxation rate at the anti-crossing is approximately 20 Hz, which is roughly a quarter of the intervalley rate measured at zero field. Taking into consideration the imprecision of the hot spot location from the fit, the hot spot relaxation rate is fairly close to the expected value of $2/T_1 \approx$ 40 Hz. Thus, the spin relaxation time at the hot spot can be used as an indirect, albeit imprecise, measurement to confirm the intervalley relaxation time.

## V. VALLEY-DEPENDENT TUNNELING

Interestingly, fitting the valley tunneling peaks to the rate equation model reveals that the tunnel rate out of the excited valley state differs from the rate out of the ground state by a factor of 4, i.e. $\Gamma_{v_+}^{out}/\Gamma_{v_-}^{out} \approx 4$. To support this discovery, we found that the read-out window for measuring valley occupation was twice the valley splitting. In the case where all tunnel rates are equal in the measurement stage, we would expect valley read-out only when the reservoir Fermi level is between valley states, and an extended read-out window is only possible when the tunneling rates associated with the two dot states are different. A likely explanation for valley-dependent tunneling is valley-orbit coupling, which under certain conditions leads to shifts in the orbital envelope functions of the two valleys [37]. This explanation is conceivable given the small valley splitting observed and the inter-dot valley coupling seen in this device [28]. To obtain a qualitative estimate of valley-dependent tunneling from valley-orbit coupling, we calculate the wavefunctions of the lowest two dot states in the presence of interface disorder using a disorder-expansion effective mass approach developed in [38]. In the presence of a single step at the Si-SiGe interface to introduce valley-orbit coupling, the excited valley wavefunction is laterally shifted with respect to the ground state and valley-dependent tunneling is expected with $\Gamma_{v_+}^{out}/\Gamma_{v_-}^{out} = 6.4$ [39]. Of course, since we can't know the atomic-scale details of the device heterostructure, the true origin of valley-orbit coupling may not be a single interface step; however, the step model allows for comparison to other theoretical studies that investigate valley-orbit coupling in quantum dots and provides a qualitative explanation for the observed effect.

## VI. SPIN AND VALLEY READ-OUT AT NONZERO FIELD

Lastly, the effects of magnetic field on intervalley relaxation are investigated. This is achieved by fixing the relative energies of the load, read and empty pulse stages and shifting them all uniformly with plunger gate voltage $V_R$. In the region above the spin-valley anticrossing where $E_z > E_{VS}$, both spin and valley tunneling peaks are observable. They can be addressed separately by setting the load level to be sufficiently small and varying $V_R$ through the charge transition. At lower dot energies where the read level straddles spin states, the spin occupation is read out. When the dot energy is increased to the point where the read level straddles valley states and the load level is below the excited spin state, valley occupation can be measured (Fig. 3(a)). Consistent with the zero-field results, the valley tunneling peak is on a larger timescale than the spin tunneling peak (Fig. 3(b)). This discrepancy can again be explained by considering the exponential dependence of dot-reservoir tunneling rates on the dot energy [15] [40]. Generally, as the levels involved in the tunneling peak get further apart, the associated tunneling rates diverge. Comparing the two tunneling peaks to the rate equation model, now including all four spin-valley states, shows good qualitative agreement [34]. Intervalley relaxation is found to be independent of field to within error bars, confirming that single-phonon processes are the dominant relaxation mechanism despite the electron temperature being comparable to the valley splitting (Fig. 3(c)). This measurement also verifies that the states involved in the relaxation process are field-independent, as the valley states should be.

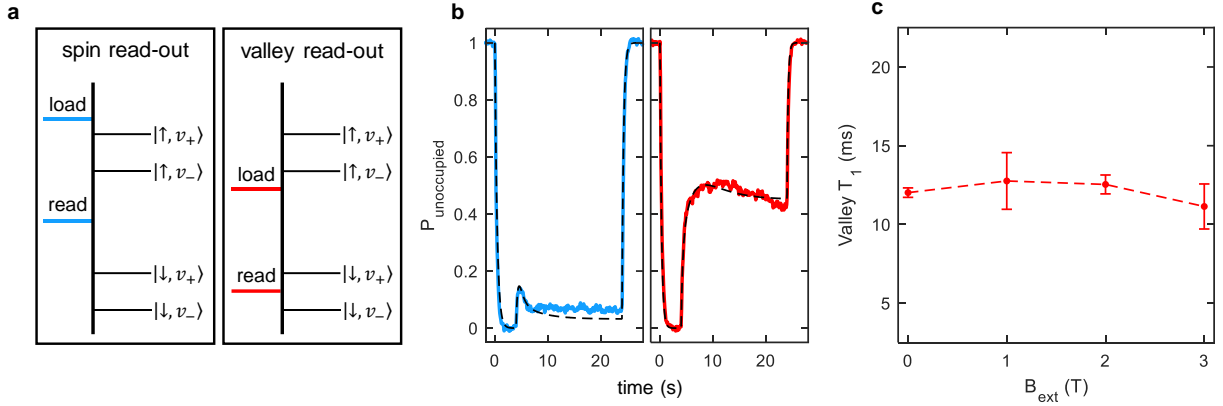

FIG. 3. Reading out spin and valley states together at nonzero magnetic field. (a) Schematic of the possible modes of state read-out when the Zeeman energy is greater than the valley splitting. Left panel: the "load" pulse level is above all relevant dot states and the "read" level is between spin states. Right panel: the "load" level is below the excited spin state and the "read" level is between valley states. (b) Example of spin-discriminating tunneling peak (left) and valley-discriminating tunneling peak (right) at $B_{ext} = 2T$, accomplished by uniformly moving all pulse stage energies. Dotted lines represent an approximate fit to a four-state rate equation model. (c) Intervalley relaxation time is shown to be independent of magnetic field to within error bars.

## VII. DISCUSSION

The valley $T_1$ in our device is 12 ms, which is long enough to have consequences for qubit operation. As we have shown, the long intervalley relaxation is constant through the magnetic field range used for a typical spin qubit. In this context, a valley relaxation time that is larger than the spin operation time (typically a few microseconds) could be detrimental, as quantum information will leak into the excited valley state and remain there during operations. On the other hand, the slow intervalley relaxation times seen here are a reflection of weak coupling to noise sources, meaning that a system with long valley $T_1$ also has long spin relaxation times. A direct measurement of real-time valley dynamics could therefore be of use for spin qubit characterization. For valley qubit implementations [41] [28] [42], small valley splittings that can be easily addressed with microwave excitations together with a long valley $T_1$ are highly desirable. The presence of valley-dependent tunneling that allows for an extended read-out window beyond $E_{\text{VS}} \approx 43$ μeV could also be a beneficial enhancement of valley qubit measurement fidelity.

VIII. CONCLUSION

We have directly observed relaxation from the excited valley state to the ground state at zero magnetic field in real time, made possible by the small valley splitting – and therefore a slow relaxation rate – in a Si/SiGe quantum dot. Due to valley-dependent tunneling, possibly arising from valley-orbit coupling, there is an extended window for reading out valley states that allows reliable measurements even when thermal energy and valley splitting are comparable. From the field independence of the relaxation, we can conclude that the valley splitting is unchanged by in-plane magnetic fields and higher-order relaxation processes are negligible up to $B_{\text{ext}} = 3$ T. The demonstrated method of intervalley relaxation measurement is as precise as single-spin $T_1$ measurements by nature, and could prove useful as a characterization tool for spin and valley qubits in silicon.

ACKNOWLEDGEMENTS

This work was supported by the U.S. Army Research Office through Grant No. W911NF1410346. N.E.P acknowledges support from the Julian Schwinger graduate fellowship at UCLA.

**Supplementary: Direct measurement of electron intervalley relaxation in a Si/SiGe quantum dot**

Nicholas E. Penthorn[1], Joshua S. Schoenfield[1], Lisa F. Edge[2], and HongWen Jiang[1]

[1]Department of Physics & Astronomy, University of California, Los Angeles, California 90095, USA

[2]HRL Laboratories, LLC, Malibu, California 90265, USA

1. **Conversion of SET current to occupation probability**

The SET channel current over the three-stage pulse period is composed of capacitive responses of the SET conductance to the pulsed plunger gate voltage and the electron tunneling in or out of the dot. With the pulse amplitudes used and proximity of the SET to the right dot, these contributions are roughly equal (Fig. S1(a)). Pure pulse-induced SET response is obtained with plunger voltage $V_R$ far enough into the $N = 1$ occupation region that electron tunneling out of the dot at any pulse stage is negligible (Fig. S1(b)). Then the plunger response data is subtracted from the SET current obtained during relaxation measurements, both averaged 2000 times, to obtain the pure electron-induced SET response, subsequently normalized so that the voltage obtained at the end of the "load" phase corresponds to an occupation probability of 1 and the voltage obtained at the end of the "empty" phase corresponds to an occupation probability of 0 (Fig. S1(c)). The normalization step is only valid for reservoir-dot tunneling rates faster than the load and empty times, otherwise the electron wouldn't be fully loaded or unloaded on average.

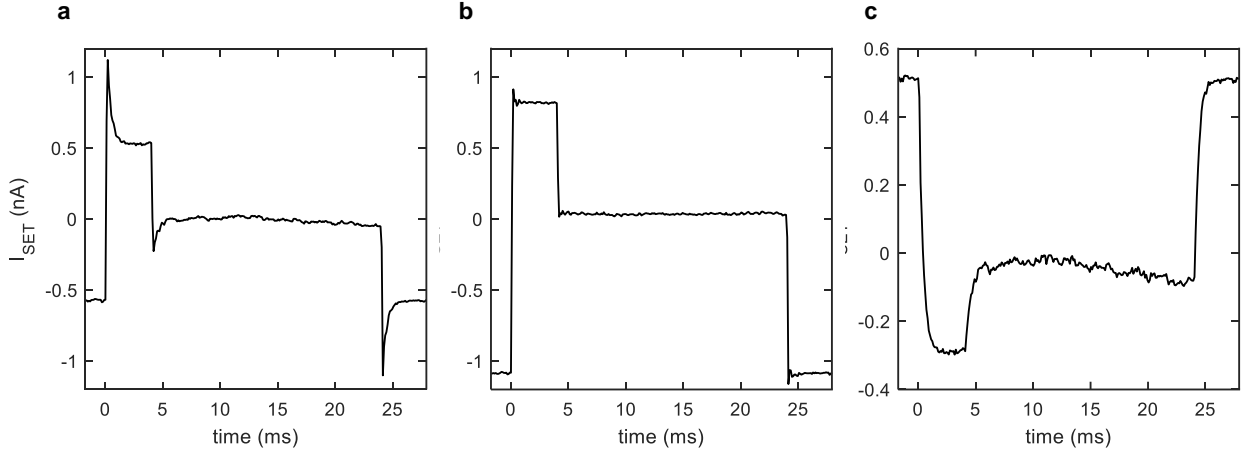

FIG. S1. Converting SET channel current into occupation probability. SET channel current $I_{\text{SET}}$ during the three-stage pulse is shown when (a) the readout level is between valley levels and (b) all pulse levels are well above both valley levels. The data in (a) contains capacitive coupling to the pulsed gate as well as the dot electron as it tunnels in and out, whereas the data in (b) only contains gate coupling. A DC offset is subtracted from all current traces due to the oscilloscope's AC coupling. (c) The result of subtracting the data in (b) from (a), isolating the coupling between the SET and dot electron.

## 2. Rate equations for tunneling peak

Because we take an ensemble average of the electron tunneling events, measured dot occupation probability can be described by a system of classical rate equations [1] [2]. At zero magnetic field when spin states are degenerate, the only available states that can be occupied during the pulse are the valley states and the probability $\boldsymbol{p} = (p_{v_+}, p_{v_-}, p_0)$ evolves as

$$\frac{d\boldsymbol{p}}{dt} = \begin{pmatrix} -\Gamma_{v_+}^{\text{out}}(E,T) - \Gamma_v & 0 & \Gamma_{v_+}^{\text{in}}(E,T) \\ \Gamma_v & -\Gamma_{v_-}^{\text{out}}(E,T) & \Gamma_{v_-}^{\text{in}}(E,T) \\ \Gamma_{v_+}^{\text{out}}(E,T) & \Gamma_{v_-}^{\text{out}}(E,T) & -\Gamma_{v_+}^{\text{in}}(E,T) - \Gamma_{v_-}^{\text{in}}(E,T) \end{pmatrix} \boldsymbol{p}. \qquad (1)$$

In (1), $\Gamma_v$ is the intervalley relaxation rate and dot-reservoir tunneling rates to state $v$ are given by

$$\Gamma_v^{\text{in}}(E,T) = \Gamma_{v,0}^{\text{in}}[1 - n(E - E_v, T)]e^{-\beta(E - E_v)}, \qquad (2)$$

$$\Gamma_v^{\text{out}}(E,T) = \Gamma_{v,0}^{\text{out}} n(E - E_v, T)e^{-\beta(E - E_v)}. \qquad (3)$$

In (2) and (3), $\Gamma_0$ is the nominal tunnel rate found when reservoir Fermi level $E$ matches valley state energy $E_v$ at zero temperature, $n(E,T)$ is the Fermi-Dirac distribution, and the exponential term with proportionality constant $\beta$ describes the dependence of the tunnel rate on the dot energy in relation to the tunneling barrier height [3]. The quantity of interest to us is $p_0$, which represents the probability of the dot to be unoccupied and is proportional to the electron-induced SET response.

A similar set of rate equations for spin-valley states can be obtained in the case that the Zeeman splitting is greater than the valley splitting, with probability $\boldsymbol{p} = (p_{\uparrow,v_+}, p_{\uparrow,v_-}, p_{\downarrow,v_+}, p_{\downarrow,v_-}, p_0)$:

$$\frac{d\boldsymbol{p}}{dt} = \begin{pmatrix} -\Gamma^{out}_{\uparrow,v_+} - \Gamma_w - \Gamma_v & 0 & 0 & 0 & \Gamma^{in}_{\uparrow,v_+} \\ \Gamma_v & -\Gamma^{out}_{\uparrow,v_-} - \Gamma_w - \Gamma_u & 0 & 0 & \Gamma^{in}_{\uparrow,v_-} \\ \Gamma_w & \Gamma_u & -\Gamma^{out}_{\downarrow,v_+} - \Gamma_v & 0 & \Gamma^{in}_{\downarrow,v_+} \\ 0 & \Gamma_w & \Gamma_v & -\Gamma^{out}_{\downarrow,v_-} & \Gamma^{in}_{\downarrow,v_-} \\ \Gamma^{out}_{\uparrow,v_+} & \Gamma^{out}_{\uparrow,v_-} & \Gamma^{out}_{\downarrow,v_+} & \Gamma^{out}_{\downarrow,v_-} & -\sum_{\sigma,v} \Gamma^{in}_{\sigma,v} \end{pmatrix} \boldsymbol{p}. \quad (4)$$

Here the functional dependence of the tunnel rates $\Gamma_{\sigma,v}(E,T)$ on energy and temperature isn't explicitly written out for the sake of brevity. Expression (4) includes both intervalley tunneling as in (1) and (intravalley) spin decay with relaxation rate given by $\Gamma_w$. For completeness, we've also introduced tunneling from $|\uparrow, v_-\rangle$ to $|\downarrow, v_+\rangle$ at a rate $\Gamma_u$; however, given that $1/t_{\text{wait}} \gg \Gamma_w > \Gamma_u$ for the range of $t_{\text{wait}}$ used in this work, $\Gamma_u$ is treated as negligibly small.

3. **Reproducing tunneling peaks**

The exponential factor $\beta$ in (2) and (3) was measured directly by varying the plunger gate voltage $V_R$ across the charge transition line while applying the three-stage readout pulse. The SET current during the "load" pulse stage decays exponentially in real-time as shown in Figure S1a, and fitting the decay gives an average tunneling rate into the dot which is in reality a sum of tunneling events into all available dot levels. The average tunneling rate as a function of gate voltage is well described by an exponential $\langle \Gamma^{in} \rangle \sim e^{-\lambda V_R}$, with $\lambda = 395$ V$^{-1}$. Given that the plunger gate lever arm is $\alpha \approx 0.15$, we arrive at $\beta \approx 6 \times 10^3$ eV$^{-1}$.

Data taken at zero external field is fit with the three rate equations in (1). From the spin relaxation hot spot (Fig. 2c in the main text) the valley splitting is approximately 43 µeV and the valley states are assigned energies $E_{v_\pm} = \pm E_{VS}/2 = \pm 21.5$ µeV. Intervalley relaxation rate $\Gamma_v$ is measured to be 83 Hz. The nominal tunneling rates used to fit Figure 2a in the main text are $\Gamma^{in}_{v_+,0} = 11.4$ kHz, $\Gamma^{in}_{v_-,0} = 2.3$ kHz, $\Gamma^{out}_{v_+,0} = 209$ Hz, and $\Gamma^{out}_{v_-,0} = 67$ Hz. It is perhaps more useful to quote the actual tunneling rates at the read-out level and operational electron temperature, which are in this case $E = -70$ µeV and $T = 200$ mK: $\Gamma^{in}_{v_+} = 97$ Hz, $\Gamma^{in}_{v_-} = 177$ Hz, $\Gamma^{out}_{v_+} = 361$ Hz, and $\Gamma^{out}_{v_-} = 85$ Hz. Clearly the actual tunneling rates are much different than the nominal rates at the read level, due to both the exponential energy dependence and Fermi-Dirac distribution terms in (2) and (3). Additionally, the dot occupation probability during the read phase is dominated by $\Gamma^{out}_{v_+}$ and a slower $\Gamma^{in}_{v_-}$, the two elements that give rise to the tunneling peak. Physically, since the read-out level is actually below both valley states, thermal excitations allow electrons to tunnel from the reservoir into the ground valley state. In Fig. S2, all probabilities encapsulated in $\vec{p}$ are plotted together to show how each state occupation evolves during the pulse sequence.

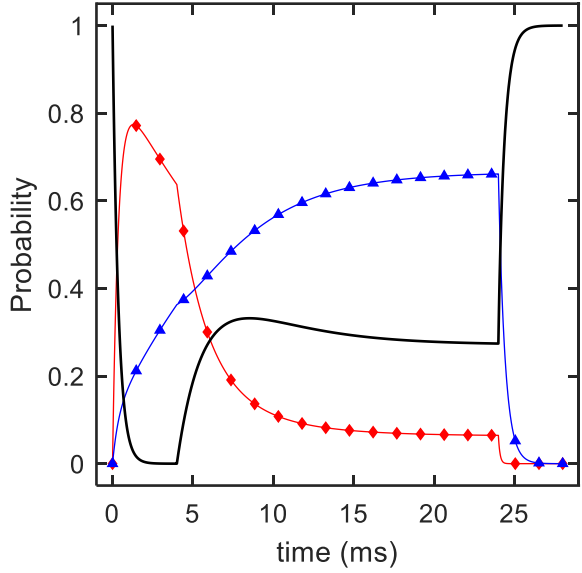

FIG. S2. Simulated valley state occupation probabilities at zero magnetic field. The experimentally measured charge sensing current is proportional to the probability of the dot being unoccupied (sold black). Electrons initially loaded into the excited valley state quickly tunnel into the reservoir (red diamonds) and tunnel back into the ground state (blue triangles).

Data taken with an external field $g\mu_B B > E_{VS}$ is fit to the five rate equations in (4). The energy levels are $E_{\uparrow,v_\pm} = \frac{1}{2}g\mu_B B \pm \frac{1}{2}E_{VS}$ and $E_{\downarrow,v_\pm} = -\frac{1}{2}g\mu_B B \pm \frac{1}{2}E_{VS}$. Nominal and actual tunneling rates used to fit the data shown in Figure 3b of the main text taken at $B = 2T$ are enumerated in Table I:

TABLE I. Tunneling rates used to fit spin and valley tunneling peaks at $B_\text{ext} = 2T$.

|  | Nominal | $E = -159$ μeV (spin read-out) | $E = -215$ μeV (valley read-out) |
|---|---|---|---|
| $\Gamma^\text{in}_{\uparrow,v_+}$ (Hz) | 1800 | 0 | 0 |
| $\Gamma^\text{out}_{\uparrow,v_+}$ (Hz) | 287 | 1697 | 2373 |
| $\Gamma^\text{in}_{\uparrow,v_-}$ (Hz) | 660 | 0 | 0 |
| $\Gamma^\text{out}_{\uparrow,v_-}$ (Hz) | 306 | 1402 | 1960 |
| $\Gamma^\text{in}_{\downarrow,v_+}$ (Hz) | 16930 | 564 | 32 |
| $\Gamma^\text{out}_{\downarrow,v_+}$ (Hz) | 180 | 261 | 373 |
| $\Gamma^\text{in}_{\downarrow,v_-}$ (Hz) | 5530 | 1371 | 95 |
| $\Gamma^\text{out}_{\downarrow,v_-}$ (Hz) | 55 | 49 | 87 |

In the valley read-out window, any electron in the excited spin state quickly tunnels out of the dot at roughly 2 kHz and back into either valley in the ground spin state. The spin tunneling peak is not visible in this case because the load level is slightly below the excited spin levels, giving an electron a 20% probability of being in the excited spin state. Then if the electron is in the excited valley it can tunnel out and tunnel back in to the true ground state at a much slower rate to give rise to the valley tunneling peak (Fig. S3(b)). In the spin read-out window, the dominant tunneling mechanism is from the excited spin states to the ground state via the reservoir, and at a longer time scale any electron in the excited valley state relaxes to the ground state (Fig. S3(a)). At the end of the read stage, the electron has a 90% probability of being in the true ground state.

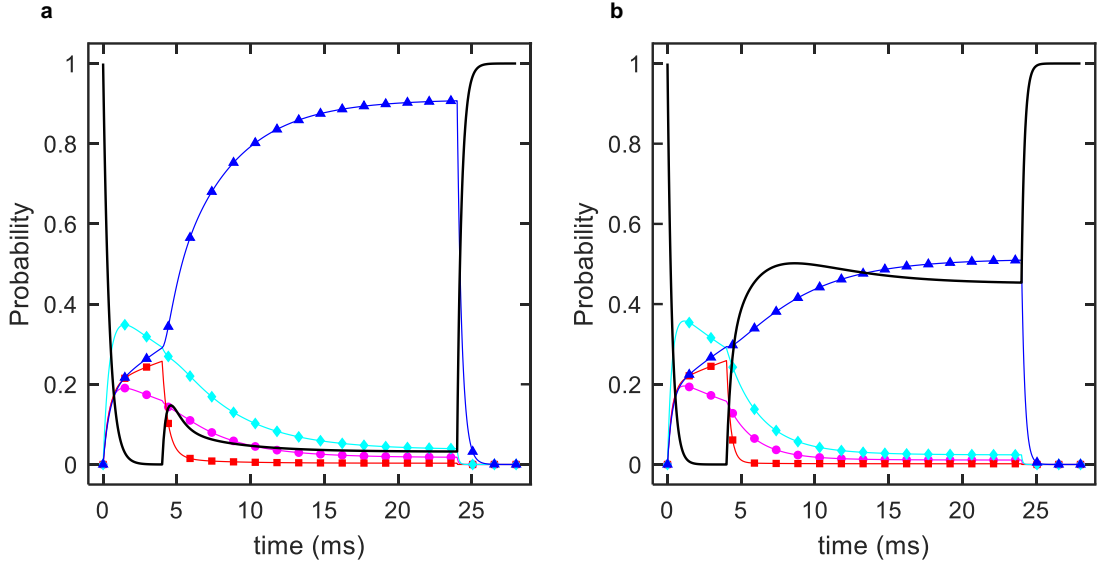

FIG. S3. Simulated electron occupation probabilities at $B_{\text{ext}} = 2\text{T}$. In (a), the read-out level is set between spin states to measure the electron spin occupation, whereas in (b) the read-out level is 80 μeV below the ground state to measure the valley occupation. Magenta circles indicate state $|\uparrow, v_+\rangle$, red squares indicate $|\uparrow, v_-\rangle$, cyan diamonds indicate $|\downarrow, v_+\rangle$ and blue triangles indicate $|\downarrow, v_-\rangle$ (the ground state). The solid black line represents the probability of the dot being unoccupied.

### 4. Valley-dependent tunneling calculation

From Table S1, the tunnel-out rates from states $|\downarrow, v_-\rangle$ and $|\downarrow, v_+\rangle$ differ by a factor of 5 in the spin read-out region and a factor of 4 in the valley read-out region. We can first rule out the contribution from energy-dependent tunneling, which is $\Gamma^{\text{out}}_{\downarrow,v_+}/\Gamma^{\text{out}}_{\downarrow,v_-} = e^{\beta E_{\text{VS}}} = 1.3$. To calculate the valley wavefunctions, we use the disorder expansion method in reference [4] and closely follow [5]. The unperturbed Hamiltonian in the $x$-$y$ plane (in the plane of the wafer) is modelled with the effective mass approximation and a harmonic quantum dot potential:

$$H_{xy} = \frac{p_x^2 + p_y^2}{2m_t} + \frac{1}{2}m_t\omega_0^2(x^2 + y^2), \tag{5}$$

where $m_t = 0.19 m_0$ is the transverse effective electron mass in silicon and $\hbar \omega_0 = 0.5$ meV.

The unperturbed Hamiltonian in the $z$ direction is given by a tight-binding model with discrete atomic sites indexed by $j$:

$$[H_z]_{jk} = t_1(\delta_{j,k+1} + \delta_{j,k-1}) + t_2(\delta_{j,k+2} + \delta_{j,k-2}) + V_{\text{SiGe}}(j)\delta_{jk} + eF_z z_j \delta_{jk}, \quad (6)$$

where $t_1 = 683$ meV and $t_2 = 612$ meV are nearest-neighbor and next-nearest-neighbor hopping amplitudes, $F_z = 3$ MV/m is the gate-induced electric field at the Si/SiGe interface, and $z_j = (j - 1/2)a_0/4$ is the real-space coordinate associated with atomic layer $j$. The potential $V_{\text{SiGe}}$ captures the 10 nm-wide square well formed by the SiGe-Si-SiGe layers:

$$V_{\text{SiGe}}(j) = \begin{cases} 0, & 0 < j \leq 74 \\ V_0, & \text{otherwise.} \end{cases} \quad (7)$$

In (7), the silicon quantum well is 74 atomic layers wide to approximate 10 nm and there are 55 atomic layers each in the SiGe buffer layers for a total of 184 atomic layers. The solutions to the full unperturbed Hamiltonian $H_{xy} + H_z$ are fully separable, with quantum harmonic oscillator solutions for the in-plane components $\psi(x, y)$ indexed by $n_x, n_y = 0, 1, \ldots$ and out-of-plane solutions capturing the fast valley oscillations, indexed by $n_z = 0, \ldots, 183$ (Fig. S4(a)).

The perturbation has the form $H_1(x, j) = \Theta(x_0 - x)\delta_{j,1}$. Matrix elements of the perturbation in the space spanned by the unperturbed eigenstates are calculated numerically and the number of basis states used in the expansion is truncated to $n_x = 6, n_y = 1, n_z = 20$. Then the full diagonal unperturbed Hamiltonian plus the expanded perturbation is diagonalized to find the perturbed eigenstates in the original unperturbed basis. The lowest two perturbed eigenstates are the valley states, shown in Fig. S4(b) and S4(c) when step termination point $x_0 = \sqrt{\hbar/m_t \omega_0}$. The excited valley state is shifted to the right from the ground state due to the step, which makes a larger portion of the excited valley envelope function extend beyond the classical boundary of the harmonic potential. Valley-dependent tunneling is then calculated by finding how much of the excited valley wavefunction penetrates the potential well boundary compared to the ground state wavefunction, which leads to $\Gamma^{out}_{\downarrow,v_+}/\Gamma^{out}_{\downarrow,v_-} = 6.4$.

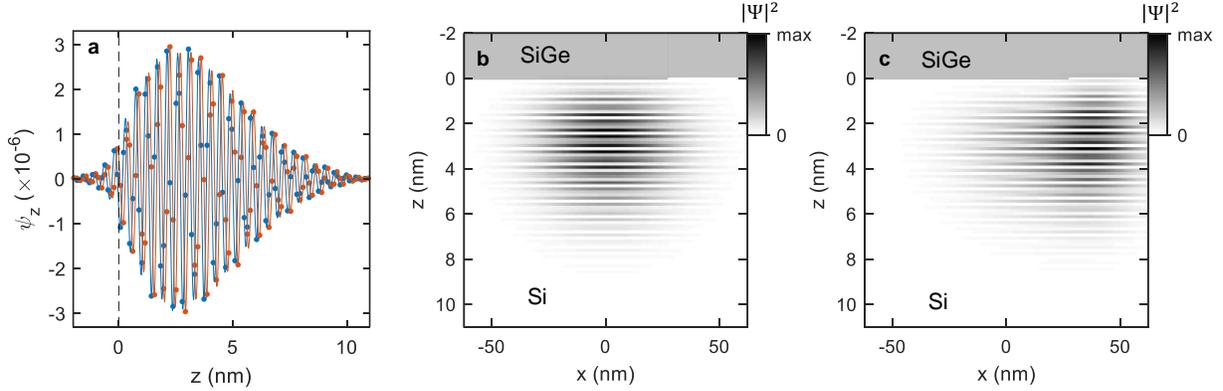

FIG. S4. Valley state wavefunctions in the presence of valley-orbit coupling. (a) Ground (blue) and excited state (orange) solutions to the unpertubed tight-binding Hamiltonian (6). (b) Full ground state wavefunction and (c) full excited state wavefunction in the presence of a Si-SiGe interface step positioned at $x_0 = 28.3$ nm. The ground state envelope is more or less unaffected by the step and the excited state is pushed to the right, which results in valley-dependent tunneling out of the quantum dot potential.